\documentclass{article}

\makeatletter 
\@addtoreset{equation}{section}
\makeatother  

\usepackage{indentfirst}
\usepackage{setspace}
\usepackage{amsmath}
\usepackage[colorlinks, linkcolor=black, anchorcolor=black, citecolor=black]{hyperref}
\usepackage{graphicx}

\title{$P-V$ Criticality In the Extended Phase Space of Charged Accelerating AdS Black Holes }
\author{Hang Liu$^{a}$\thanks{E-mails:hangliu@mail.nankai.edu.cn} and Xin-he Meng$^{a,b}$\thanks{E-mails:xhm@nankai.edu.cn}
\\
\\
$^{a}$ School of Physics, Nankai University, Tianjin 300071, China\\
$^{b}$ State Key Laboratory of Theoretical Physics, Institute of Theoretical Physics,\\
Chinese Academy of Science, Beijing 100190, China}
\date{}
\begin{document}
\large
\maketitle
\begin{abstract}
In this paper, we investigate the $P-V$ criticality and phase transition of charged accelerating AdS black holes in the extended thermodynamic phase space in analogy between black hole system and van der Waals liquid-gas system, where  the cosmological constant $\Lambda$ is treated as a thermodynamical variable interpreted as dynamic pressure and its conjugate quantity is the thermodynamic volume of the black holes. When the electric charge vanishes, we find that no $P-V$ criticality will appear but the Hawking-Page like phase transition will be present, just as what Schwarzschild-AdS black holes behave like. For the charged case, the $P-V$ criticality appears and the accelerating black holes will undergo a small black hole/large phase transition under the condition that the acceleration parameter $A$ and the horizon radius $r_h$ meet a certain simple relation $A r_h=a$, where $a$ is a constant in our discussion. To make $P-V$ criticality appear, there exists an upper bounds for constant $a$.  When $P-V$ criticality appears, we calculate the critical pressure $P_{c}$, critical temperature $T_{c}$ and critical specific volume $r_{c}$, and we find that $\frac{P_{c}r_{c}}{T_{c}}$ is an universal number.
\end{abstract}
\tableofcontents
\section{Introduction}
Since Hawking and Bekenstein \textit{et al.} had established black hole thermodynamics in their pioneer work by Hawking's radiation with black body spectrum and in analogy between the black hole mechanics with the classical thermodynamics, a lot of attention has been attracted to the study the thermodynamic property of  black holes, especially after Hawking and Page finished their seminal paper \cite{Hawking1} which demonstrates that there exist a phase transition in the phase space between Schwarzschild-AdS black hole and thermal radiation, i.e., the so-called Hawking-Page phase transition which can be explained as the confinement/deconfinement phase transition of gauge field in the AdS/CFT correspondence. And then, Authors in Ref.\cite{Cvetic1,Cvetic2} extend our understanding of phase transition and critical phenomena to other more complicated backgrounds. The thermodynamic property of black holes in the AdS space is much different from those in the de Sitter space or asymptotically  flat spacetime. The black holes in the asymptotically flat spacetime have negative specific heat in many cases which means that this kind of black hole is unstable, e.g., a Schwarzschild black hole loses mass through Hawking radiation, becoming hotter(due to the negative specific heat) and eventually evaporating away. However, in the AdS space, the large black holes(larger than the order of the AdS radius) have positive specific heat and are thermodynamically stable, such black holes will get colder as they lose mass. In this sense, the AdS space can play a role of ``box" where the black holes can exist stably. In other words, the radiation of black holes and the ``box" can reach a thermal equilibrium. The authors in Ref.\cite{David} have studied the phase transition and critical behavior of charged AdS(RN-AdS) black holes in extended phase space by analogy between the charged AdS black holes and van der Waals liquid-gas system. By this analogy, they find that the charged AdS black holes can undergo critical behavior and phase transition between large black holes and small black holes, just the same as the van der Waals liquid-gas system if the horizon radius $r_h$ is regarded as the specific volume $v$.
They have also calculated the critical exponents  and showed that  the charged AdS black holes share the same critical exponents with the van der Waals system.
In the extended phase space, the cosmological constant is treated as a thermodynamical variable related to the dynamic pressure $P$ by the relation

\begin{equation*}
P=-\frac{\Lambda}{8\pi}
\end{equation*}
\\
where $\Lambda$ is the cosmological constant and the conjugate quantity of pressure $P$ is explained  as the thermodynamic volume of the black holes. With this identification, it is reasonable to include the cosmological constant in the first law and the Smarr relation could be obtained by scaling argument which cannot deduce Smarr relation if the cosmological constant is not present. It is worth noting that  the black hole mass $M$ should be explained as enthalpy rather than internal energy of the system in the extended phase space.

So far, a lot of  black holes in the AdS space have been studied widely in literatures \cite{Cao,Cai1,David,Banerjee2012,Banerjee2012a,Gunasekaran2012,Hendi2016,Wang1,Wang2,Mirza2014} about their thermodynamic properties and people have made a great progress in understanding the nature of  black hole thermodynamics, though some more fundamental  problems are still not utterly solved, such as understanding the microscopic origin of black hole entropy in quantum gravity, \textit{etc.}. Nevertheless, the thermodynamics of accelerating AdS black holes described by the so called $C-metric$ and their generalizations \cite{Kin1,Ple1,Dias1,Gri1} is rarely studied. For accelerating black holes, one would expect the local temperatures associated to corresponding horizons to be different leading to a problem of thermodynamic equilibrium due to the existence of the second horizon, i.e., accelerating horizon. To get ride of this effect, one can consider a negative cosmological constant which can ``remove" the accelerating horizon. Such a black hole is called the slowly accelerating and is displaced a little from the center of negatively curved spacetime at the cost of employing a force in the form of cosmic string ending on the horizon \cite{Podo1}.
Therefor, it is interesting to investigate the thermodynamical properties of such an accelerating charged AdS black hole and see whether the $P-V$ criticality and phase transition appears or not. On the other hand, the previous lectures \cite{David,Gunasekaran2012} have showed that the electric charge plays an important role in the appearance of a large/small black hole phase transition, and we want to know if the electric charge is still essential in this picture. Further more, the effects of the acceleration parameter $A$ on black hole thermodynamical properties are also worth investing. These are key motivations of the present paper.

When it comes to the thermodynamics of an accelerating AdS black hole, we are supposed to note that the object we are considering is accelerating and such accelerating object always carries with it the notion of time-dependent, though the C-metric is not time-dependent. So one natural question is that how can an accelerating system be in equilibrium? Recently, the authors in Ref.\cite{Michael} have answered that question.
They successfully formulate and investigate thermodynamics of these slowly accelerating black holes. The first law of these slowly accelerating black holes can be expressed in standard form and the present paper is based on their work in Ref.\cite{Michael}.

This paper is organized as follows: In section 2, we will introduce the thermodynamics of charged accelerating AdS black holes by reviewing Ref.\cite{Michael}. In section 3, we study the $P-V$ criticality and phase transition. In section 3.1, we investigate the Gibbs free energy of non-charged accelerating AdS black holes for a better understanding of the phase transition and critical behavior, and following naturally to discuss the Gibbs free energy in charged case in section 3.2. The last section is devoted to conclusions and discussions.

\section{Thermodynamics of charged accelerating black holes}
In this section, we give a brief  introduction to the thermodynamics of charged accelerating AdS black holes by reviewing Ref.\cite{Michael}. The metric of a charged accelerating AdS black hole can be expressed as \cite{Gri1,Michael}

\begin{equation}
ds^2=\frac{1}{\Omega^2}\left[f(r)dt^2-\frac{dr^2}{f(r)}-r^2\left(\frac{d\theta^2}{g(\theta)}+g(\theta)\sin^2\theta\frac{d\phi^2}{K^2}\right) \right]
\end{equation}
where
\begin{subequations}
\begin{gather}
f(r)=(1-A^2r^2)(1-\frac{2m}{r}+\frac{e^2}{r^2}+\frac{r^2}{l^2})\\
g(\theta)=1+2mA\cos\theta+e^2A^2\cos^2\theta
\end{gather}
\end{subequations}
\\
and for the conformal factor $\Omega$, we have

\begin{equation}
\Omega=1+Ar\cos\theta
\end{equation}
\\
which determines the conformal infinity of the AdS space. The parameters $m$ and $e$ are related to black hole mass and electric charge respectively, the parameter $A$ is related to the magnitude of acceleration of the black hole and $l=\sqrt{-\Lambda/3}$ is the AdS radius. By looking at the angular part of the metric and the behavior of $g(\theta)$ at the both poles $\theta_{+}=0$ and $\theta_{-}=\pi$, we can discover the presence of cosmic string. The regularity of the metric at a pole requires

\begin{equation}
K_{\pm}=g(\theta_{\pm})=1\pm 2Am+e^2A^2
\end{equation}
\\
and $K$ is chosen to regularise one pole and the another pole is left either a conical deficit or a conical excess along the other pole. Under the consideration that a conical excess can be sourced be a negative energy object, we would make the black hole is regular on the north pole, i.e., $\theta=0$, by fixing $K=K_{+}$, which is

\begin{equation}
K=1+2Am+e^2A^2
\end{equation}

The method of conformal completion \cite{Ashtekar2,Das1}, which takes the electric part of the Weyl tensor projected along the time-like conformal Killing vector $\partial_{t}$, and integrates over a sphere at conformal infinity, can be used to identify the black hole mass $M$, electric charge $Q$ and electrostatic potential $\Phi$, which are given by

\begin{gather}
\begin{split}
Q&=\frac{1}{4\pi}\int_{\Omega=0}*F=\frac{e}{K}\\
\Phi&=\frac{e}{r_{h}}\\
M&=\frac{m}{K}
\end{split}
\end{gather}
\\
where $r_{h}$ represents black hole horizon radius and $F$ is the electromagnetic field tensor which is related to gauge potential $B$

\begin{equation}
F=dB, \quad B=-\frac{e}{r}dt
\end{equation}

The horizon area is denoted as $\mathcal{A}$

\begin{equation}
\mathcal{A}=\int_0^\pi\int_0^{2\pi}\sqrt{g_{\theta\theta}g_{\phi\phi}}d\theta d\phi=\frac{4\pi r_h^2}{K(1-A^2r_h^2)}
\end{equation}
\\
the black hole entropy $S$ is identified with a quarter of the horizon area

\begin{equation}
S=\frac{\mathcal{A}}{4}=\frac{\pi r_h^2}{K(1-A^2r_h^2)}\label{S}
\end{equation}
\\
The Hawking temperature is given by

\begin{equation}
T=\frac{f'(r_h)}{4\pi}=\frac{m}{2\pi r_h^2}-\frac{e^2}{2\pi r_h^3}+\frac{A^2 m}{2\pi}+\frac{r_h}{2\pi l^2}\label{T}
\end{equation}
\\
where we have denoted the derivative of $f(r)$ with respect to $r$ as $f'(r)$. The pressure $P$ is associated to the cosmological constant according to

\begin{equation}
P=-\frac{\Lambda}{8\pi}\label{P}=\frac{3}{8l^2}
\end{equation}
\\
By using E.q (\ref{P}), E.q (\ref{T}) and E.q (\ref{S}), we can obtain

\begin{equation}
TS=\frac{M}{2}-\frac{\Phi Q}{2}+\frac{4\pi}{3K}\frac{r_h^3}{(1-A^2r_h^2)^2}P\label{TS}
\end{equation}
\\
If we regard the black hole thermodynamics volume as

\begin{equation}
V=\frac{\partial M}{\partial P}\Big|_{S,Q}=\frac{4\pi}{3K}\frac{r_h^3}{(1-A^2r_h^2)^2}
\end{equation}
\\
the E.q (\ref{TS}) yields

\begin{equation}
M=2(TS-PV)+\Phi Q
\end{equation}
\\
which is nothing but the Smarr relation \cite{Smarr,Kastor1}, and the first law can be expressed in the standard form

\begin{equation}
dM=TdS+\Phi dQ+VdP\label{first}
\end{equation}

Further more, the slowly accelerating black holes also satisfy the reverse isoperimetric inequality \cite{Cvetic3} which holds for AdS black holes. The isoperimetric inequality says that, in  Euclidean space, the volume $V$ enclosed in a given area $\mathcal{A}$ is maximised for a spherical surface, or conversely for a given volume, the sphere owns the minimum area. While for AdS black holes in $n$ dimensions, the reverse isoperimetric inequality is satisfied which can be expressed mathematically via isoperimetric ratio

\begin{equation}
\mathcal{R}=\left[\frac{(n-1)V}{\omega_{n-2}}\right]^{\frac{1}{n-1}}\left(\frac{\omega_{n-2}}{\mathcal{A}}\right)^{\frac{1}{n-2}}\geq 1
\end{equation}
\\
where $\omega_{n-2}$ denotes the volume of the unit $(n-2)$-sphere, and the equality only holds for ordinary spherical black holes,  which indicates that the  spherical black holes  possess maximum area. Note that the horizon area $\mathcal{A}$ is related to entropy, so the reverse isoperimetric inequality suggests that the entropy of spherical black holes is maximised at given thermodynamic volume $V$. For this accelerating black hole, we have $\omega_{2}=4\pi/K$ and $n=4$. Using the formulas above for $V$ and $\mathcal{A}$, one can obtain

\begin{equation}
\mathcal{R}=\left(\frac{3V}{\omega_{2}}\right)^{\frac{1}{3}}\left(\frac{\omega_{2}}{\mathcal{A}}\right)^{\frac{1}{2}}
=\frac{1}{(1-A^2r_h^2)^{\frac{1}{6}}}\geq 1
\end{equation}
\\
Thus we find that the slowly accelerating black holes do indeed satisfy the reverse isoperimetric inequality.

\section{\texorpdfstring{$P-V$}{} criticality and phase transition}
In this section, we will discuss the $P-V$ criticality and phase transition of charged accelerating AdS black holes by investigating the Gibbs free energy of the black hole system, and we would like to begin with the calculation of critical pressure $P_c$, critical specific volume $r_c$ and critical temperature $T_c$. 

From the horizon function

\begin{equation}
f(r_h)=0
\end{equation}
\\
we can obtain the mass parameter

\begin{equation}
m=\frac{3r_h^2+(8\pi P-3A^2)r_h^4+3e^2(1-A^2r_h^2)}{6r_h(1-A^2r_h^2)}\label{m}
\end{equation}
\\
By employing E.q (\ref{m}), we can compute the Hawking temperature on the horizon

\begin{equation}
T=\frac{f'(r_h)}{4\pi}=\frac{-3r_h^2+6(A^2-4\pi P)r_h^4+(8A^2\pi P-3A^4)r_h^6+3e^2(A^2r_h^2-1)^2}{12\pi r_h^3(A^2r_h^2-1)}\label{TT}
\end{equation}
\\
 E.q (\ref{TT}) yields

\begin{equation}
P=\frac{3(A^2r_h^2-1)\{(r^2-e^2)(A^2r_h^2-1)+4\pi r_h^3T\}}{8\pi r_h^4(A^2r_h^2-3)}\label{PP}
\end{equation}

Note that the temperature $T$ will approach infinity when the horizon radius $r_h \rightarrow \frac{1}{A}$. To avoid this singularity, we would like to take the value of $Ar_h$ as a constant for simplicity, i.e.,

\begin{equation}
A=\frac{a}{r_h}\label{Ar}
\end{equation}
\\
where $a$ is a constant which does not equal one. The critical point is determined as the inflection point in the $P-V$ diagram by the following equations

\begin{gather}
\frac{\partial P}{\partial r_h}\Big |_{r_h=r_c,T=T_C}=0\label{Pr}\\
\frac{\partial^2 P}{\partial r_h^2}\Big |_{r_h=r_c,T=T_C}=0\label{Pr2}
\end{gather}
\\
where we have treated the horizon radius $r_h$ as the specific volume $v$, i.e., $v=r_h$. By considering condition (\ref{Ar}) and solving equations (\ref{PP}), (\ref{Pr}), (\ref{Pr2}), we can get the critical pressure $P_c$, critical specific volume $r_c$ and critical temperature $T_c$

\begin{gather}
T_c=\frac{1-a^2}{3\sqrt{6}e\pi}\label{Tc}\\
r_c=\sqrt{6}e\label{rc}\\
P_c=\frac{(a^2-1)^2}{32(3-a^2)e^2 \pi}\label{Pc}
\end{gather}
\\
and then we find an interesting  relation between $P_c$, $T_c$ and $r_c$

\begin{equation}
\frac{P_c r_c}{T_c}=\frac{9(a^2-1)}{16(a^2-3)}\label{PRT}
\end{equation}
\\
which is an universal number and only depends on the value of the constant $a$ we take. This result is every similar to that in the van der Waals system, the only difference is that for the van der Waals system, the right hand side of E.q (\ref{PRT}) is $3/8$. When we take the constant $a$ as 0, i.e., the acceleration parameter $A=0$ and the black hole metric just goes back to the RN-AdS black hole case, and the value of E.q (\ref{PRT}) becomes

\begin{equation}
\frac{P_c r_c}{T_c}=\frac{3}{16}
\end{equation}
\\
if we take relation between specific volume $v$ and horizon radius $r_h$ as $v=2r_h$, we can get

\begin{equation}
\frac{P_c r_c}{T_c}=\frac{3}{8}
\end{equation}
\\
which is exactly the same as the case in the van der Waals system and RN-AdS black holes \cite{David}. In this sense, it is natural to regard the horizon radius $r_h$ as specific volume rather than thermodynamic volume.
From E.q (\ref{Tc}), we can see that we must let the constant $-1< a<1$ to keep the critical temperature $T_c$ positive. For acceleration parameter $A$, we take it as a positive parameter and the horizon radius $r_h$ is also positive so that the constant $a$ is restricted to

\begin{equation}
0\leq a<1
\end{equation}

\begin{figure}[th]
\centering
\includegraphics[width=4.0in]{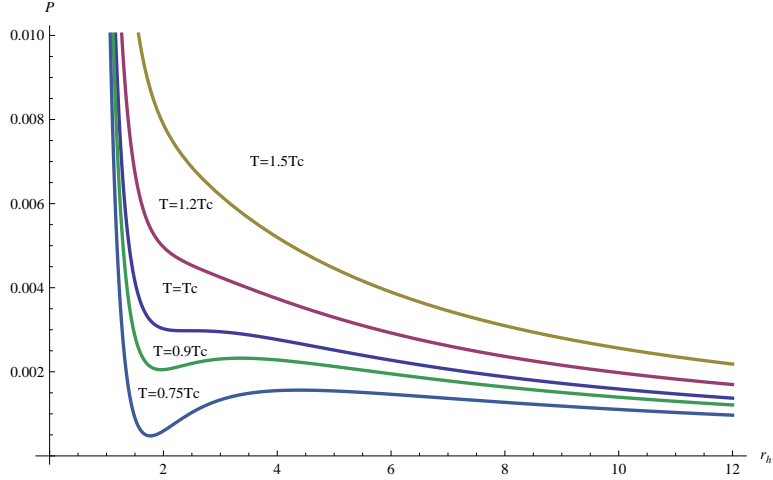}
\caption{The $ P-V(r_h)$ diagram of charged accelerating AdS black holes, here we take charge $e=1$, constant $a=1/4$. For upper isothermals with temperature $T>T_c$, there only exists one branch with positive compression coefficient, thus representing stable state and no phase transition. For the lower isothermals with $T<Tc$, there exists 3 branches corresponding to 2 stable state and 1 unstable state, phase transition happens. \label{fig1}}
\end{figure}

We have drawn $P-V$ diagram for the charged accelerating AdS black holes in Fig.(\ref{fig1}) which is exactly the same as that in the van der Waals liquid-gas system. We plot 5 $P-r_h$ curves under the fixed temperature. For the two upper isothermals with temperature $T>T_c$, the curves have only one branch with a positive compression coefficient, thus there only exists one stable state,i.e., gas state and no phase transition could occur. For the two lower isothermals with temperature $T<Tc$, we can see that these two curves have 3 branches, two of them have positive compression coefficient corresponding two stable phase in two horizon radius ranges, the small radius range stands for small black hole and the large radius stands for large black hole. The branch between those two branches has a negative compression coefficient representing unstable phase where the gas phase and liquid phase can coexist in the van der Waals gas-liquid system. Therefor a phase transition can occur between a small black hole and a large black hole when temperature is lower than critical temperature $T_c$, while for temperature $T>T_c$, only gas phase can exist and no phase transition occurs.

\subsection{Gibbs free energy in the case with electric charge \texorpdfstring{$e=0$}{}}
To have a better understanding of phase transition, we investigate the behavior of the Gibbs free energy, which is defined as

\begin{equation}
G=U-TS+PV=H-TS
\end{equation}
\\
where $U$ is the internal energy of the thermodynamics system, and $H$ represents enthalpy. From the first law (\ref{first}) in the extended phase space, we can see that the black hole mass $M$ plays a role of enthalpy instead of internal energy, thus the Gibbs free energy for the black hole is

\begin{equation}
G=M-TS
\end{equation}

\begin{figure}[th]
\centering
\includegraphics[width=3.0in]{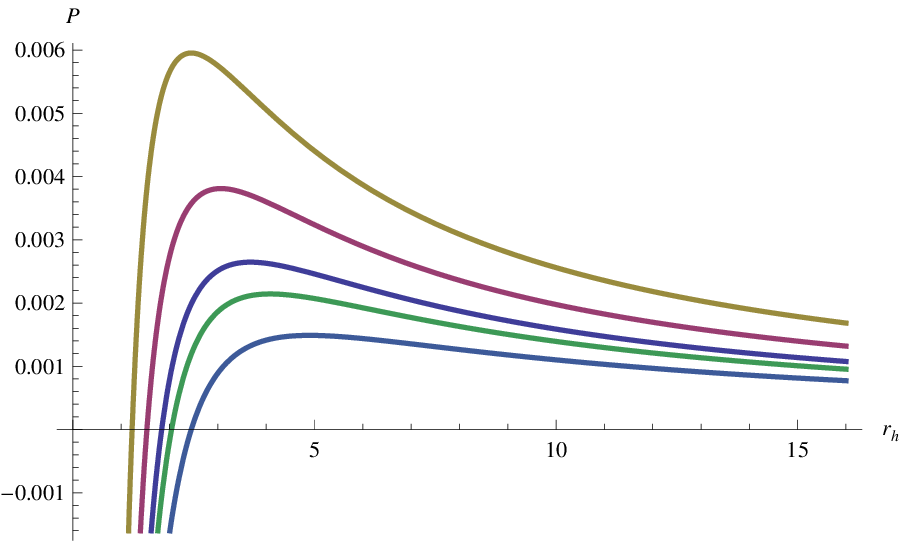}~~
\includegraphics[width=3.0in]{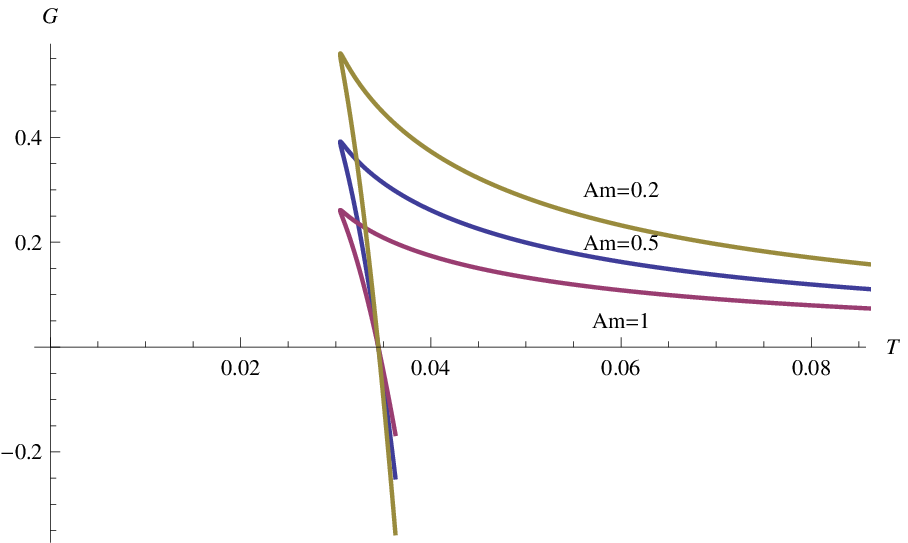}
\caption{The $ P-V(r_h)$ diagram(left plot) and Gibbs free energy as a function of temperature at fixed pressure for different $Am$ of non-charged accelerating AdS black holes. Here we take $e=0$ and $a=1/4$.\label{fig2}}
\end{figure}

We plot $P-V$ diagram and Gibbs free energy of non-charged accelerating AdS black holes at 5 different fixed temperatures in Fig.(\ref{fig2}). From the $P-V$ diagram (left plot), we find that the isothermals have only two branches at different fixed temperatures, one branch with negative compression coefficient representing unstable phase and another branch has positive compression coefficient corresponding to stable phase such that no phase transition happens. From another perspective, the equations (\ref{Pr}) and (\ref{Pr2}) determining the critical point has no solutions, therefor no critical behavior occurs in the charge $e=0$ case. While from the diagram of Gibbs free energy plotted at fixed pressure, it is obviously to see that there exists a Hawking-Page like phase transition,i.e., the transition between thermal radiation and black holes. We plot Gibbs free energy as a function of temperature by varying cosmic string tension represented by $Am$. There is a different lowest temperature the black hole can exist for different cosmic string represented by $Am$, and we can see from the plot that the higher the value of $Am$ is, the lower the lowest temperature in which the black holes can exist is.  As expected, the black holes on the upper branch of the Gibbs free energy function curve have negative specific heat and are thermodynamic unstable , and those on the lower branch have positive specific heat representing thermodynamic stable black holes.

\begin{figure}[th]
\centering
\includegraphics[width=3in,height=2.5in]{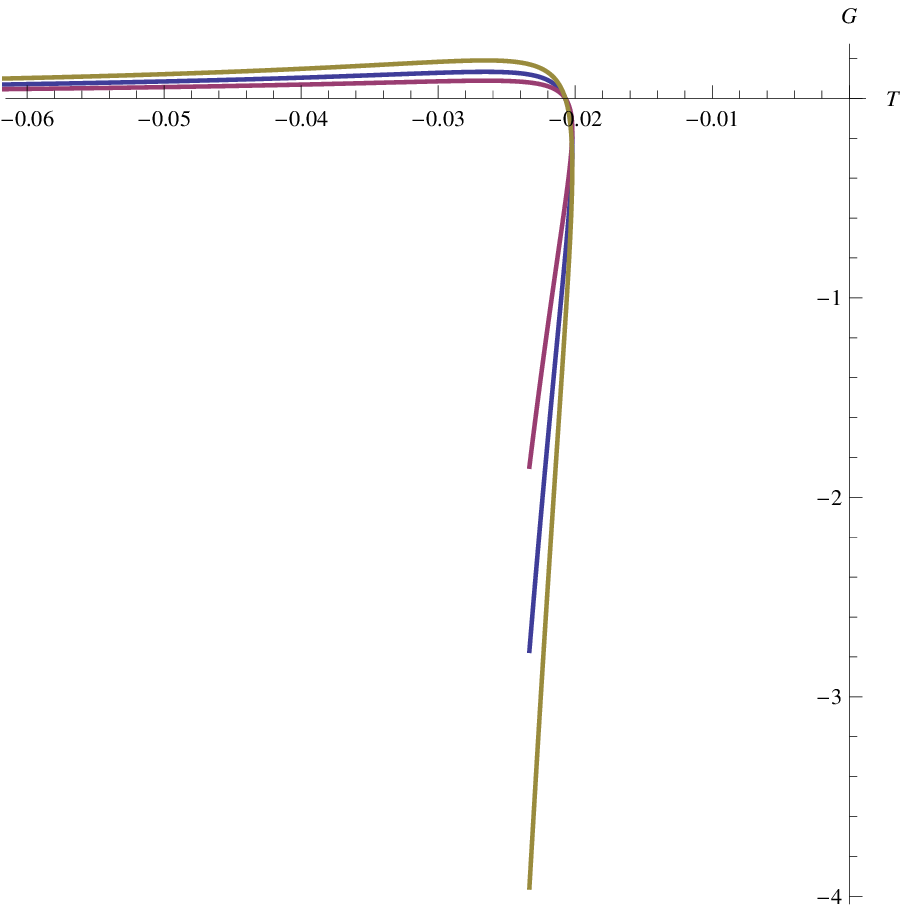}~~
\includegraphics[width=3in,height=2.5in]{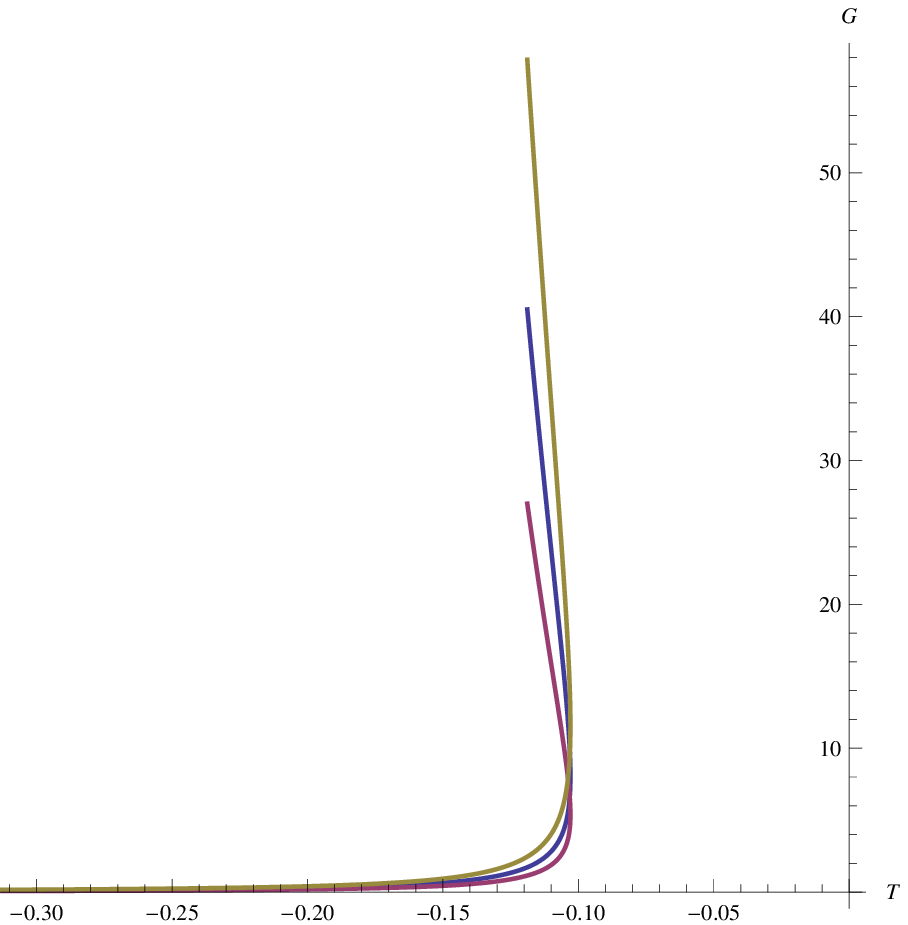}~~
\caption{The plot of Gibbs free energy for non-charged accelerating AdS black holes in the case $a=1.2$(left plot),  $a=1.8$(right plot) and $e=1$ for $Am=0.2$, $0.5$ and $1$. The red line corresponds to $Am=1$, the blue line corresponds to $Am=0.5$ and the orange line corresponds to $Am=0.2$.\label{fig3}}
\end{figure}

We also draw the Gibbs tree energy at fixed pressure in the case that the constant $a>1$ and $Am=0.2$, $0.5$ and $1$, respectively in Fig.(\ref{fig3}). The left plot is related to $a=1.2$ and the right plot is related to $a=1.8$. The both curves show that  the black hole can only exist at the temperature $T<0$. For this negative temperature, we believe that it is  unphysical and should  be abandoned such that no black hole can exist in the case $a>1$.

\subsection{Gibbs free energy  in the case with electric charge \texorpdfstring{$e\neq 0$}{}}
\begin{figure}[thbp]
\centering
\includegraphics[height=2in,width=2.6in]{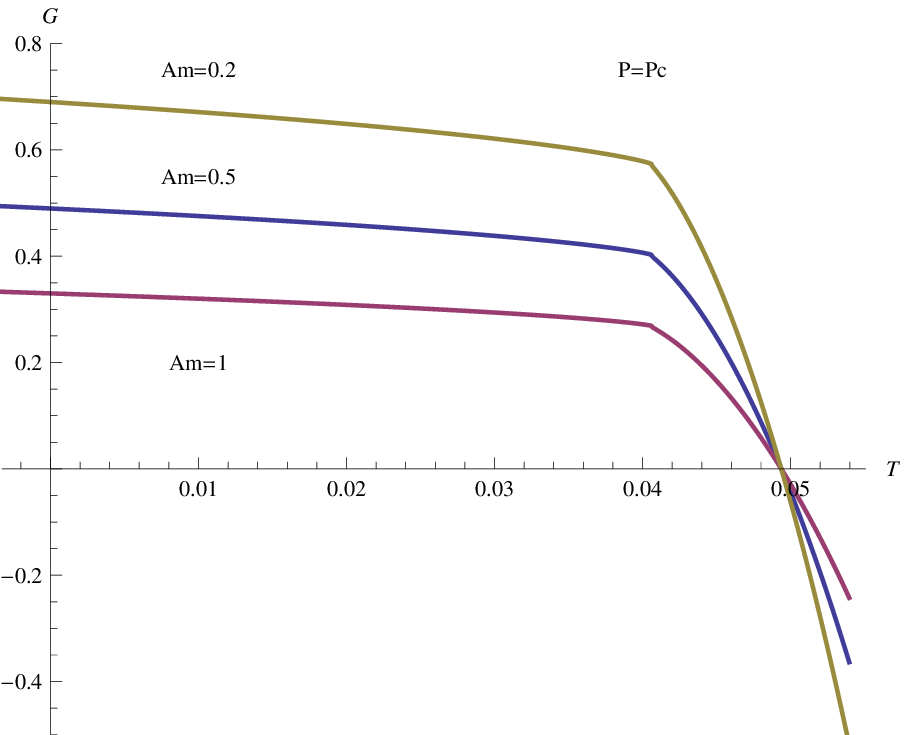}~~
\includegraphics[height=2in,width=2.6in]{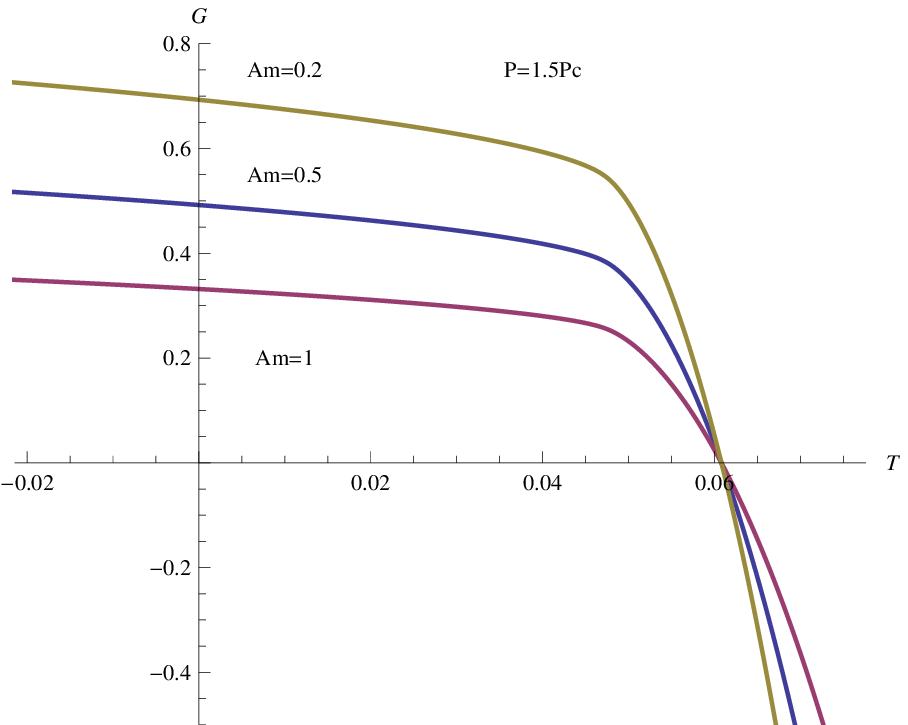}

\includegraphics[height=2in,width=2.6in]{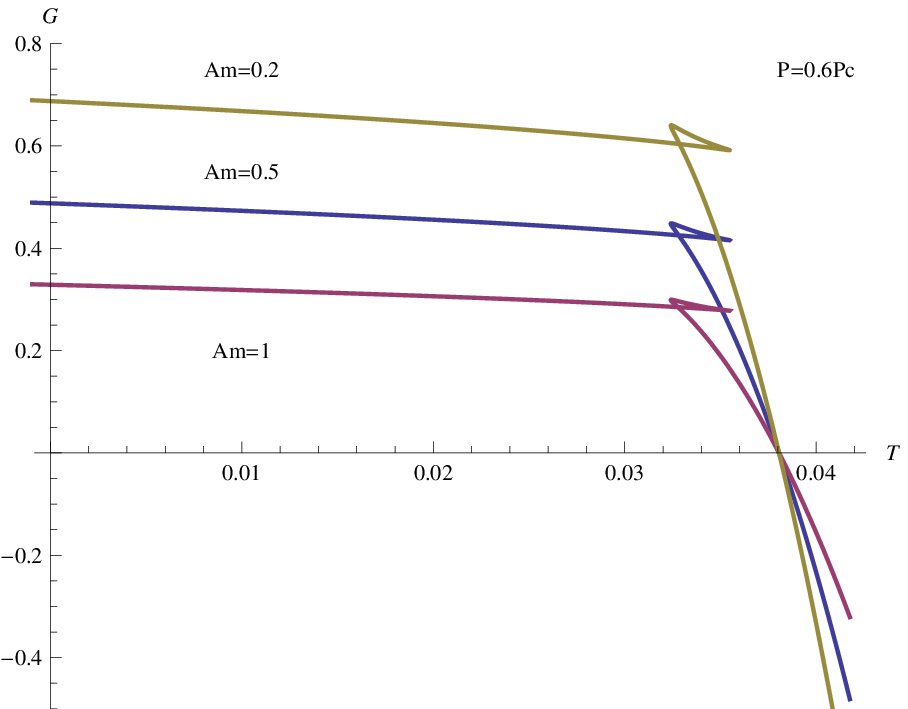}~~
\includegraphics[height=2in,width=2.6in]{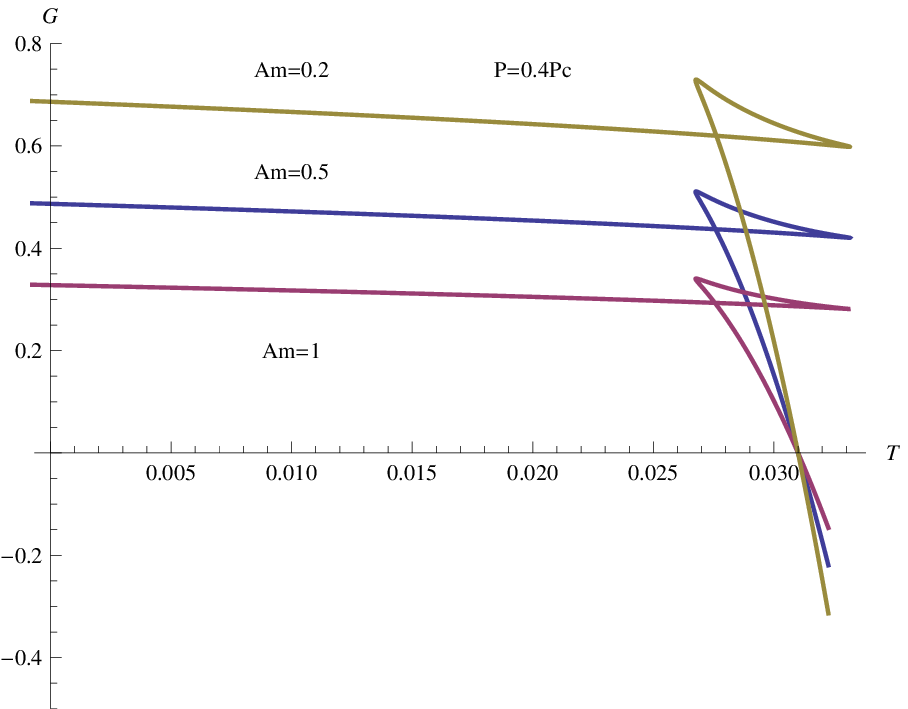}
\caption{The plots of Gibbs free energy for charged accelerating AdS black holes at four different pressures with charge $e=1$, constant $a=1/4$. It is obviously to see, for $T<T_c\approx 0.0406092$ and $P<P_c$, there is a small/large black hole phase transition indicated by the characteristic swallowtail \label{fig4}}
\end{figure}

In this section, we investigate the behavior of Gibbs free energy for charged accelerating AdS black holes. From E.q(\ref{Tc}), (\ref{Pc}) and (\ref{rc}) and $P-V$ diagram in Fig.(\ref{fig1}), we know that in the charged case there exist a critical pressure $P_c$ below which  a phase transition can happen. We draw four Gibbs free energy plots at four different fixed pressures in Fig.(\ref{fig4}). As expected, when pressure $P<P_c$, we observe the characteristic swallowtail behavior indicating the small/large black hole phase transition, as shown in  Fig.(\ref{fig4}). At $P=P_c$, the swallowtail disappears, corresponding to critical point.

\begin{figure}[thbp]
\centering
\includegraphics[height=2in,width=2.6in]{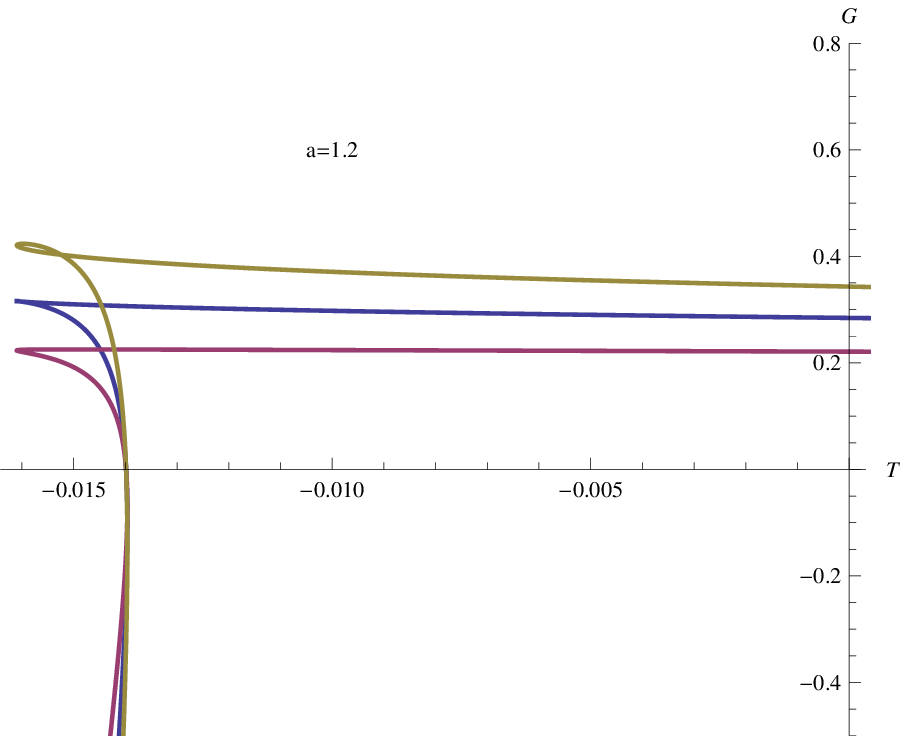}~~
\includegraphics[height=2in,width=2.6in]{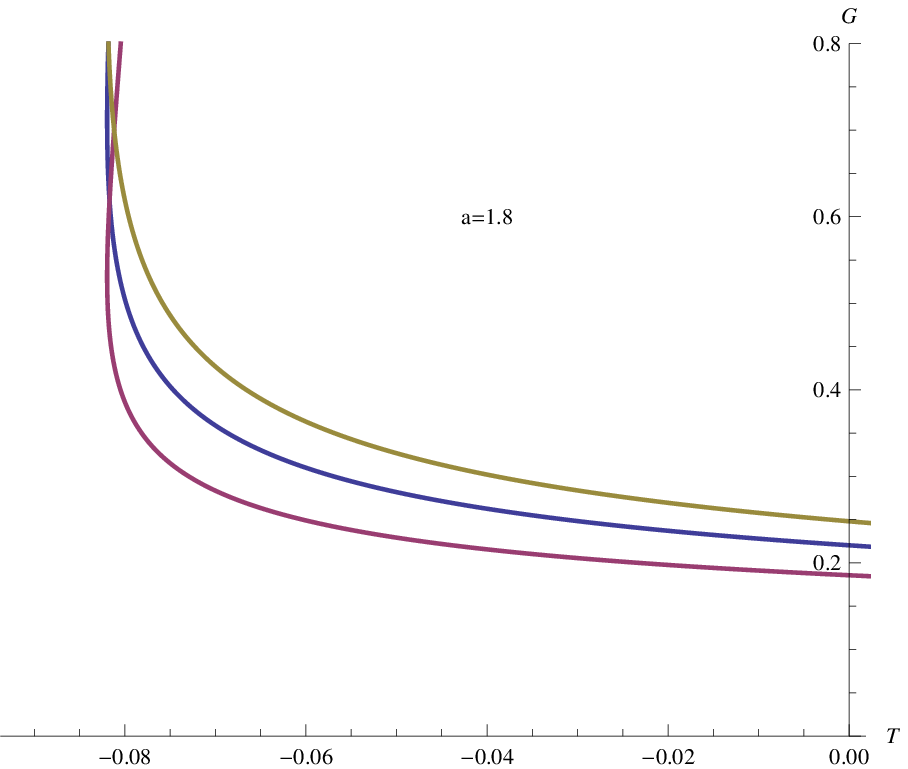}
\caption{The plots of Gibbs free energy for charged accelerating AdS black holes in the case that constant $a=1.2$(left plot), $a=1.8$(right plot) and $e=1$ at their respective $P=0.5P_c$. The red line corresponds to $Am=1$, the blue line corresponds to $Am=0.5$ and the orange line corresponds to $Am=0.2$.\label{fig5}}
\end{figure}

We also draw the Gibbs free energy in the case that the constant $a=1.2$ and $a=1.8$ at their corresponding temperature $P=0.5P_c$ in Fig.(\ref{fig5}). The curve in the right plot of Fig.(\ref{fig5}) is just normal and shows that there is  no phase transition characteristic, therefor no  phase transition happens. For the left plot in Fig.(\ref{fig5}), we can see this plot is highly similar to that in Fig.(\ref{fig2}) indicating a Hawking-Page phase transition, while this plot does not indicate any phase transition since we are supposed to focus on the positive temperature part which does not have phase transition characteristic  in the the plot. We can conclude that the charged(non-charged) accelerating AdS black holes can not undergo a phase transition if $Ar_h=a>1$.

\section{Conclusions and Discussions}
In this paper, we have investigated the $P-V$ criticality and phase transition of charged accelerating AdS black holes in the extended thermodynamic phase space in analogy between black hole system and van der Waals liquid-gas system, where  the cosmological constant $\Lambda$ is treated as a thermodynamical variable interpreted as dynamic pressure and its conjugate quantity is the thermodynamic volume of the black holes. When the electric charge vanishes, we find that no $P-V$ criticality will appear but the Hawking-Page like phase transition will be present, just as what Schwarzschild-AdS black holes behave like. For the charged case, the $P-V$ criticality appears and the accelerating black holes will undergo a small black hole/large black hole phase transition under the condition that the acceleration parameter $A$ and the horizon radius $r_h$ meet a certain simple relation $A r_h=a$, where $a$ is a constant in our discussion. We find that when $a>1$, there will be no phase transition happening. To observe a phase transition, the constant $a$ is restricted to $0\leq a<1$ and when $a=0$, we just go back to RN-AdS black hole case.  When $P-V$ criticality appears, we calculate the critical pressure $P_{c}$, critical temperature $T_{c}$ and critical specific volume $r_{c}$, and we find that $\frac{P_{c}r_{c}}{T_{c}}$ is an universal number which only depends on the constant $a$ we take. If we have $a=0$ and relation between specific and horizon radius $v=2r_h$, that universal number we get is exactly the same as the one in the van der Waals liquid-gas system, i.e., $3/8$ which  suggests that the gravity system has a  profound relation to the thermodynamical system.

However, as the statement in Ref.\cite{Michael} that there is no possibility to have a phase transition between a pure radiation in AdS space and accelerating black holes since the existence of a conical singularity that extends to the AdS boundary in accelerating spacetime, which means that phase transition and critical behavior we discuss in this paper will not occur, even for the large black hole/small black hole phase transition in charged case, but at least our paper has shown that the accelerating AdS black holes are very similar to their non-accelerating cousins from a thermodynamical point, which may give us a better understanding of the accelerating black hole thermodynamics.

\section*{Acknowledgements}
For the present work, Hang Liu would like to thank  Wei Xu and Zhen-Ming Xu for helpful discussions. This project is partially supported by NSFC

\bibliography{TABH}
\end{document}